# Critical effect of cubic phase on aging in 3 mol% yttria-stabilized zirconia ceramics for hip replacement prosthesis

Jérôme Chevalier, Sylvain Deville, Etienne Münch, Romain Jullian, Frédéric Lair

Department of Research into the Metallurgy and Physical Properties of Materials, National Institute of Applied Sciences, Associate Research Unit 5510, 69621 Villeurbanne Cedex, France

**Abstract**

The isothermal tetragonal-to-monoclinic transformation of 3Y-TZP ceramics sintered at two different temperatures (1450°C and 1550°C) and duration (2 and 5 h) is investigated at 134°C in steam. Particular attention is paid to the presence of a cubic phase and its effect on isothermal aging. Sintering at 1550°C can result in a significant amount of large cubic grains in the specimens, that have a detrimental impact on aging resistance, especially for the first stage of the aging process. Cubic grains appear to be enriched in yttrium, which in turn leads to a depletion of yttrium in the neighboring tetragonal grains. These grains will act as nucleation sites for tetragonal-to-monoclinic transformation. Even for specimens sintered at lower temperature, i.e. 1450°C, the presence of a cubic phase is expected from the phase diagram, leading to a significant effect on aging sensitivity.

## 1. Introduction

The concept of stress-induced phase transformation in zirconia ceramics represents one of the most remarkable innovations in the ceramic field. Indeed, it was shown in the 1970s, firstly by Garvie et al. [1] and then by Gupta et al. [2], that zirconia exhibits a transformation toughening mechanism, which increases its crack propagation resistance. The stress-induced phase transformation involves the transformation of metastable tetragonal grains to the monoclinic phase at the crack tip, which, accompanied by a volume expansion, induces compressive stresses. Yttria-stabilized zirconia ceramics, usually called Y-TZP, belong to the family of toughened materials. They can exhibit a strength of more than 1 GPa with a toughness of about 6–10 MPa.m$^{1/2}$. One of the most successful applications of yttria–tetragonal zirconia polycrystal (Y-TZP) ceramics so far is found in orthopedics, with femoral heads for total hip replacement. The use of zirconia has opened the way towards new implant designs that were not possible with alumina, more brittle. It has been estimated that about 500 000 patients have been implanted with zirconia femoral heads since 1985, 200 000 for these past 5 years.

Nevertheless, alarming problems related to aging of zirconia have been recently reported. In particular, the resistance to steam sterilization and the hydrothermal stability of yttria-containing zirconia in the body have been questioned. The phenomenon of zirconia aging is known in the ceramic community for more than 20 years, from the work of Kobayashi et al. [3] who discovered a serious limitation of Y-TZP ceramics for applications near 250°C. Aging occurs by a slow tetragonal-to-monoclinic phase transformation of grains on any surface in contact with water [4], or body fluid. This transformation leads to surface roughening [5], grain pull out and micro-cracking [6]. Aging was thought to be very limited and well controlled near the ambient temperature. However, the US Food and Drug Administration (FDA) have recently reported on the critical effect of the well-developed steam sterilization procedure (134°C, 2-bars pressure) on zirconia implants [7]. This procedure is now forbidden for zirconia. More important, in 2001, the FDA and the Australian Therapeutic Goods Administration (TGA) [8,9] announced that firms distributing orthopedic implants were recalling series of Y-TZP hip prostheses due to a fracture episode of zirconia ceramic femoral heads. The failures origin could be related to an accelerated tetragonal-to-monoclinic phase transformation of zirconia on a limited number of batches [10]. These failures, even if limited in number, have a strong negative impact for the use of zirconia in orthopedics.

There is today a trend for certain surgeons to go back to less-competitive solutions, abandoned several years ago. Alternative approaches are also under consideration, for example zirconia-toughened alumina ceramics [11,12]. However, it should be kept in mind that zirconia has been implanted successfully for more than 15 years with an exceptionally low failure rate [13]. A pragmatic approach could consider a better understanding of the key mechanisms of zirconia aging and the relationships between processing, microstructure and aging resistance. In this respect, the aim of this paper is to investigate on the possible detrimental role of the appearance of a cubic phase for sintering 3Y-TZP (namely 3 mol % yttria–tetragonal zirconia polycrystal) ceramics. Our work on this issue started with the evidence that 100% monoclinic phase was never reached in our experiments on 3Y-TZP even after very long exposure in steam [5]. This was in agreement with previous experimental works (for example in [14]), but not with a common feeling that 3Y-TZP is fully tetragonal after sintering, which would mean that all grains would transform after aging to the monoclinic structure after long-term aging.

A careful reading of the Y2O3-ZrO2 phase diagram [15] shows, however, that 3Y-TZP at sintering temperature (typically in the range between 1400°C and 1550°C) should present two phases : an yttrium-reach cubic phase and a tetragonal phase with an amount of yttrium lower than the nominal 3 mol % (see Fig. 1). This was experimentally demonstrated by Ruiz and Readey [16], and more recently by Matsui et al. [17] who showed that 3Y-TZP heated in the range 1450–1600°C (i.e. temperatures used for processing most of zirconia parts) exhibited a dual phase microstructure, consisting of both cubic and tetragonal polymorphs. Depending on sintering temperature and duration, cubic grains exhibiting 6–7 mol % yttria could coexist with tetragonal grains containing less than 2 mol % yttria. Thermal treatments at high temperatures were sometimes proposed to obtain tough and strong yttria-stabilized compounds [16,18] with a large transformation toughening capability. However, aging occurs by a slow transformation of metastable tetragonal grains at the surface in contact with body fluids, by a nucleation and growth mechanism [5], and nucleation is likely to arise at grains less stable than the average. Therefore, the hydrothermal stability of tetragonal grains in equilibrium with a cubic phase has to be questioned.

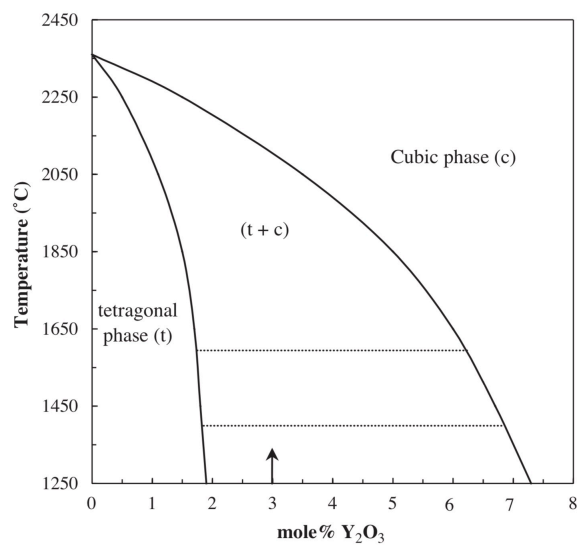

Fig. 1. Selected region of the $ZrO_2$–$Y_2O_3$ equilibrium phase diagram (after Scott [15]). The dashed lines show the typical range of sintering temperatures of 3 mol % yttria-stabilized zirconia.

## 2. Materials and methods

The experiments were conducted from a starting ultra-pure Y-TZP powder containing 3 mol % (5.2 wt%) yttria, produced by a co-precipitation method (Tosoh). The material was processed by cold isostatic pressing at 300 MPa, followed by pressureless sintering at a temperature of 1450°C or 1550°C, for 2 or 5 h. Small cylinders of 10 mm diameter and 5 mm thickness were ground and mirror-polished to reach a sufficiently low surface roughness to observe any small surface uplift due to t–m transformation and to begin aging experiments with a monoclinic content close to 0%. Monoclinic content was measured by an X-ray diffraction (XRD) technique (CuK$_\alpha$ radiation, with a penetration depth in zirconia estimated to be about 5 μm from Ref. [19]) and calculated from the modified Garvie and Nicholson equation [20]. Aging experiments were carried out in steam at 134°C under 2-bars pressure, which corresponds to a sterilization procedure. The t–m transformation being thermally activated (with an activation energy of 106 kJ/mol [5]), it can be calculated that 1 h of this treatment corresponds roughly to 4 years in vivo [21]. This treatment is therefore a good indicator of the aging sensitivity in vivo of a given zirconia ceramic. Aging was followed by XRD and by atomic force microscopy (AFM) (D3100, Digital Instruments Inc.) to follow topographic changes induced by the transformation. The AFM was used in contact mode, with an oxide-sharpened silicon nitride probe exhibiting a tip radius of curvature of 20 nm. The scans were performed at the surface with an average scanning speed of 10 μm/s.

Some specimens were thermally etched at 1300°C for 30 min so as to reveal the microstructure. They were observed by scanning electron microscopy (SEM) (FEI, ESEM, XL30) equipped with an energy dispersion spectroscopy (EDS) system (FEI, EDXi). The specimens were gold-coated to avoid any surface charging effect. EDS spectra were conducted in specific regions for a 30 kV voltage, for 5 min, to obtain well-resolved Y–K$_\alpha$ and Zr–K$_\alpha$ peaks. Grain size data were obtained from representative SEM micrographs using the ASTM E112 standard.

## 3. Results

### 3.1. Microstructures

All specimens exceeded 98% of theoretical density. Fig. 2 (a)–(d) show representative SEM micrographs of polished and thermally etched 3Y-TZP specimens heat treated at 1450–1550°C for 2–5 h. The corresponding grain sizes are given in Table 1. The grain size is very small and homogeneous for specimens sintered at 1450°C for 2 h. It becomes larger for specimen sintered at 1450°C for 5 h or 1550°C for 2 h but still unimodal. However, few coarser grains (~1 μm) could be observed in some occasions. The microstructure becomes clearly bimodal when sintering is achieved at 1550°C, 5 h, with the occurrence of grains larger than 2 μm. Transmission Electron Microscopy (TEM) by Rhüle et al. [22] has revealed that these grains were cubic in nature. This has been confirmed recently by Ruiz and Readey [16], who showed the evidence of these cubic grains above 1500°C. According to the phase diagram [15] (see Fig. 1), these grains should exhibit larger amount of yttria than tetragonal grains. EDS spectra were therefore conducted on one specimen sintered at 1550°C for 5 h on specific regions of the surface, i.e. at the interior of presumed cubic grains (points denoted A and B in Fig. 2d), in presumed tetragonal regions at the periphery of these cubic grains (i.e. point denoted C in Fig. 2d) and in one region situated far from large (cubic) grains (denoted D in Fig. 2d). The results were analyzed in terms of mol % of yttria to compare with the starting powder (3 mol %), in Table 2. Cubic grains exhibit a large yttria content, while tetragonal grains at their periphery are depleted in yttria. Far from these regions, the yttria content is that of the starting powder, within the error range of the measurement method.

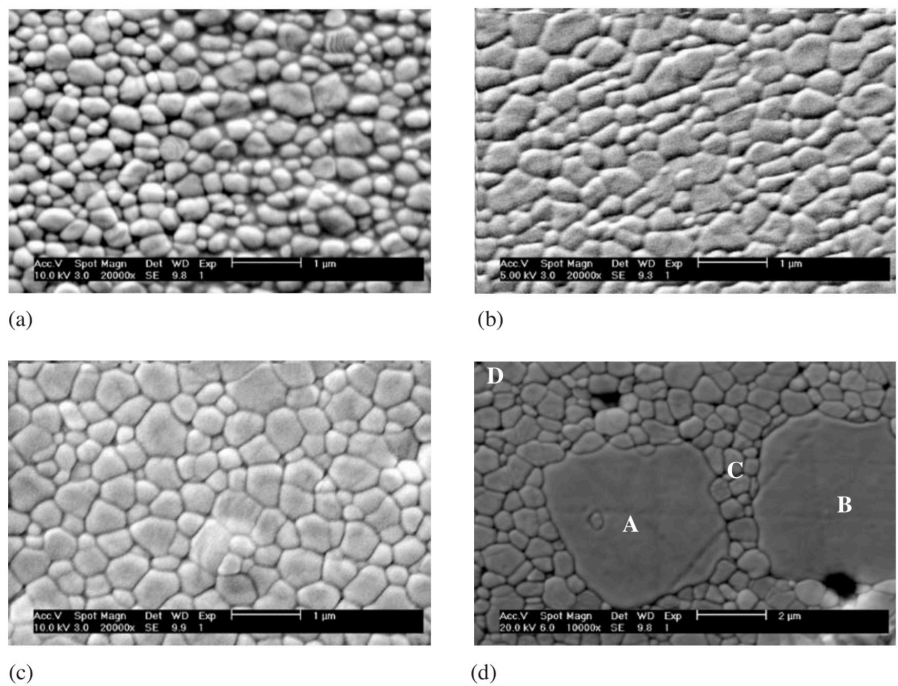

Fig. 2. Scanning electron microscopy images of 3 mol % Y-TZP ceramics: (a) sintered at 1450°C for 2 h, (b) sintered at 1450°C for 5 h, (c) sintered at 1550°C for 2 h, (d) sintered at 1550°C for 5 h. Points denoted from A to D are regions where EDS spectra were performed.

Table 1. Average grain size of 3Y-TZP ceramics as a function of sintering temperature and duration

| Sintering condition | 1450°C, 2 h | 1450°C, 5 h | 1550°C, 2 h | 1550°C, 5 h |
|---|---|---|---|---|
| Average grain size (±0.1 μm) | 0.4 μm | 0.6 μm | 0.6 μm | 0.9 μm (t) 4 μm (c) |

Table 2. Yttria content in some selected regions of Fig. 2

| Region | A (center of a cubic grain) | B (center of a cubic grain) | D (tetragonal grains at the periphery of cubic grains) | E (tetragonal grains far from cubic grains) |
|---|---|---|---|---|
| % yttria (±0.5 mol %) | 5.5 mol % | 5.0 mol % | 2.3 mol % | 3.3 mol % |

## 3.2. Low temperature aging kinetics

The relationship between the amount of monoclinic phase and aging time for the four materials is plotted in Fig. 3. The shape of the four kinetics is basically the same, but the transformation occurs at different rates: the higher the temperature and dwell duration, the higher the transformation rate. It could be related to the overall grain size [14,23]: the larger the mean size of the tetragonal grains, the lower their stability. However, the grain size of tetragonal grains of the material sintered at 1450°C for 5 h and 1550°C for 2 h is about the same, the only difference lying on the possibility to form a larger portion of cubic grains in the latter material. These grains attract yttrium from neighboring tetragonal grains which in turn must be less stable. The difference in transformation rates is particularly important for the first stage of the transformation, where nucleation of the new monoclinic phase is predominant [5]. The larger the portion of cubic phase present in the material (theoretically, from the phase diagram and experimentally from SEM and EDS results), the larger the nucleation rate of phase transformation.

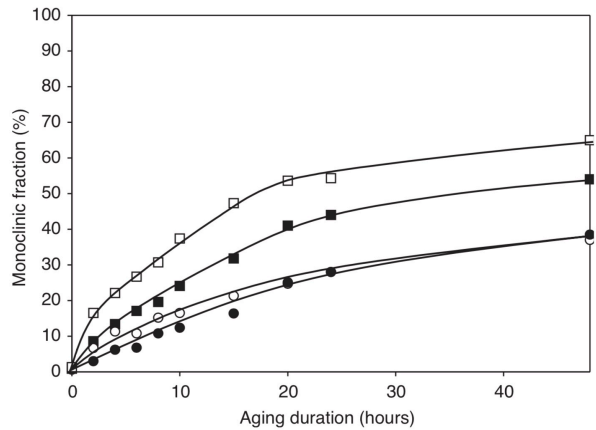

Fig. 3. Aging kinetics, at 134°C under 2-bars pressure, of 3 mol % Y-TZP ceramics: (●) sintered at 1450°C for 2 h, (○) sintered at 1450°C for 5 h, (■) sintered at 1550°C for 2 h, ( ) sintered at 1550°C for 5 h.

## 3.3. AFM and SEM observations of transformation

In order to demonstrate the ability of cubic grains to destabilize neighboring tetragonal grains, AFM observations were performed in the material sintered at 1550°C for 5 h, at the beginning of the aging process (after 1 h of treatment), and then when the monoclinic content measured by XRD was almost constant (after 24 h of treatment). As suggested before, aging obviously initiates preferentially in regions surrounding cubic grains. This is illustrated in Fig. 4, where the same zone is observed before aging (Fig. 4a) and after 1 h autoclave (Fig. 4b). The line scan conducted at the periphery of a cubic grain after 1 h aging provides confirmation of t–m transformation in this region (Fig. 4c). After a longer autoclave treatment, cubic grains appeared to be unaffected by the aging (they are in their stable structure at the ambient temperature), while all surrounding area were completely transformed (see Fig. 5, taken for the same zone before aging and after 24 h autoclave). This result is critical, since it shows that cubic grains act as nucleation sites for transformation because surrounding tetragonal grains exhibit less yttria than initially in the raw powder. By comparison, the initiation of aging occurred at a much lower rate in the material processed at 1450°C for 5 h. After 24 h of treatment, the surface appears homogeneously transformed (Fig. 6). Some isolated slightly larger grains seem to be, however, not transformed (i.e. points denoted A and B in Fig. 6 for example). These grains could be cubic in nature, since a significant fraction of cubic phase is also expected from the phase diagram.

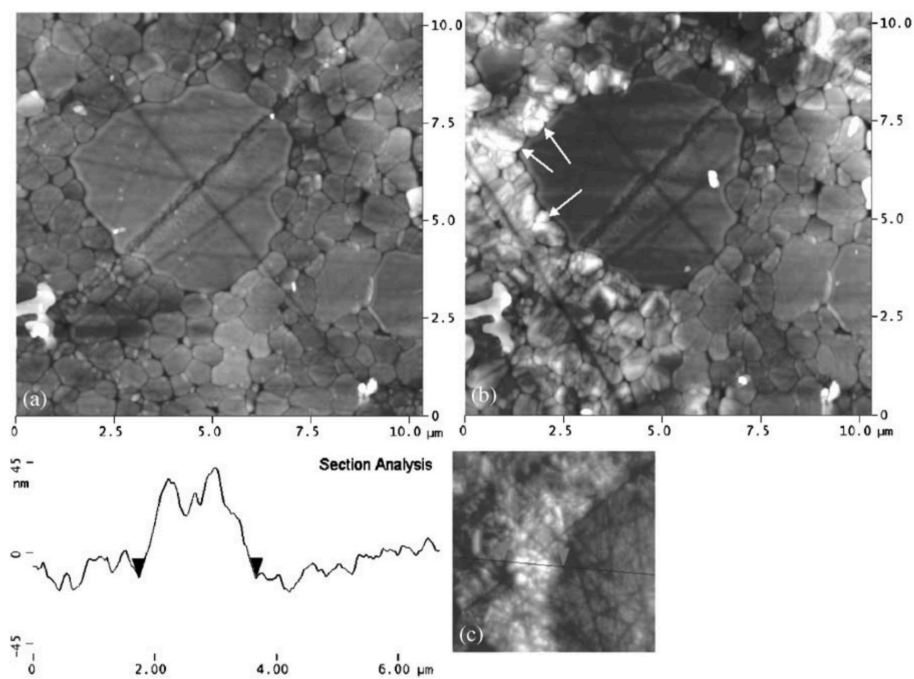

Fig. 4. Atomic force microscopy images of the same zone, (a) before and (b) after 1 h aging at 134°C, for a 3Y–TZP ceramic sintered at 1550°C for 5 h. Arrows indicate surface uplifts due to the t–m transformation of tetragonal grains. (c) Corresponds to a line scan at the periphery of a cubic grains after 1 h aging.

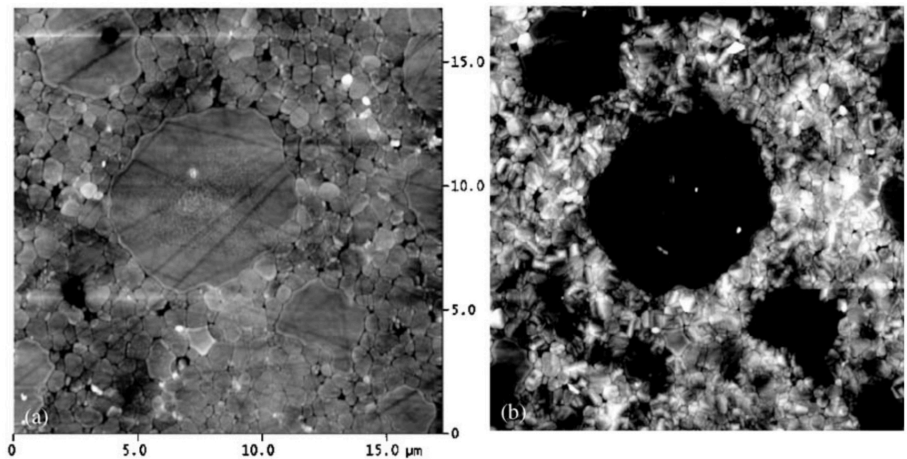

Fig. 5. Atomic force microscopy images of the same zone, (a) before and (b) after 24 h aging at 134°C, for a 3Y-TZP ceramic sintered at 1550°C for 5 h.

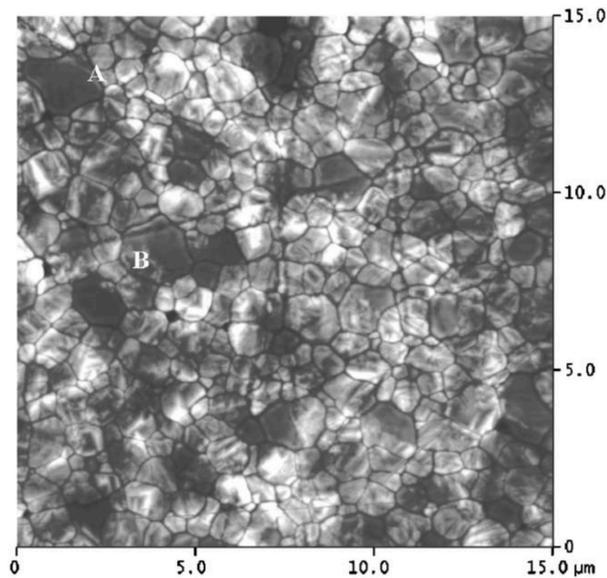

Fig. 6. Atomic force microscopy images of a 3Y-TZP ceramic sintered at 1450°C for 5 h, after 24 h aging at 134°C.

These results are confirmed by SEM observations (Fig. 7), which reveal after aging a very important surface degradation (with a large amount of grain pull-out and some micro-cracking) in the material processed at 1550°C for 5 h, especially at the periphery of cubic grains. Micro-cracking in the cubic grains occurs since large stresses are induced by the transformation of the tetragonal neighboring grains, with no option to accommodate them.

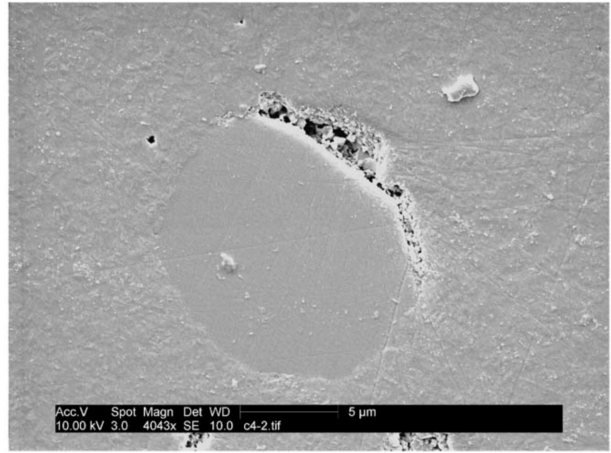

Fig. 7. Scanning electron micrograph of a 3Y-TZP ceramic sintered at 1550°C for 5 h, after 24 h aging at 134°C.

## 4. Discussion

The negative impact of cubic grains on the resistance of zirconia to steam sterilization or aging in vivo has not been considered so far. The present results confirm that 3Y-TZP zirconia ceramics sintered above 1450°C can exhibit a dual microstructure, formed with cubic and tetragonal grains. Here we show that cubic grains are enriched with yttrium, causing a depletion of yttria in the surrounding tetragonal grains. Preceding works have shown that transformation occurs by a nucleation and growth mechanism [5]. Transformation is thought to arise first for grains less stable than the average, propagating then to the neighbors due to large stresses and micro-cracks induced by the transformation. The cubic grains can therefore act as nucleation sites for the transformation. This was detrimental for the zirconia ceramics sintered at 1550°C in this study. However, even for specimens sintered at lower temperature, i.e. 1450°C, the presence of a cubic phase is expected from the phase diagram, and evidence for some non-transformed grains after aging was provided by AFM. These grains can have a significant effect on aging. It is surprising that no attention was paid to this aspect in the literature so far, while it was suggested from a simple observation of the equilibrium diagram.

We do not claim that this effect can explain the accelerated aging of some specific batches of zirconia ceramics in the recent failure events. However, it shows that processing can play a role which was not always totally clarified in the past. Further work on the relation between processing–microstructure–aging resistance should be put forward. It should be kept in mind that tetragonal-to-monoclinic transformation occurs by a nucleation and growth mechanism and that nucleation arises at grains which are less stable than the average. This means that the presence of cubic grains, but also of residual stresses, scratches or residual pores should play a role on aging of a given zirconia sample.

## 5. Conclusion

By considering that a cubic phase should be present in yttria-stabilized zirconia ceramics after sintering, the impact of cubic grains on aging was studied. Cubic grains were identified for materials sintered at 1550°C; they were shown to be enriched by yttrium, which in turn leads to a decrease of the yttrium content in the neighboring tetragonal grains. These grains will act as the further nucleation sites for tetragonal-to-monoclinic transformation. In this sense, the presence of cubic grains has a detrimental impact on aging resistance. The processing of 3Y-TZP should be conducted at sufficiently low temperatures to avoid the occurrence of a dual cubic–tetragonal microstructure, but sufficiently high temperatures to obtain fully dense materials. This means that a narrow range of sintering temperature, certainly between 1400°C and 1450°C, has to be chosen to process aging resistant 3Y-TZP. The role of post sintering treatments, such as hot isostatic pressing (HIP), conducted at higher temperatures, but smaller duration, should be studied carefully.

## Acknowledgments

Part of this work was conducted during a student project in the Materials Science and Engineering department ('Science et Génie des Matériaux') at INSA. The authors are grateful to the Clyme (Consortium Lyonnais de Microscopie Electronique) for providing the microscopy facilities.## References